\documentclass[letterpaper, 10 pt, conference]{ieeeconf}
\IEEEoverridecommandlockouts

\usepackage[utf8]{inputenc}
\usepackage{amsmath}
\usepackage{mathrsfs}
\usepackage{amssymb}
\usepackage{color}
\usepackage{hhline}
\usepackage{dsfont}
\usepackage{cite}
\usepackage{comment}
\usepackage{float}
\usepackage{graphicx}
\usepackage[font = footnotesize]{caption}
\usepackage[font = footnotesize]{subcaption}
\usepackage{bernsteinStyle}
\usepackage{float}
\usepackage{tikz}
\usepackage{booktabs}
\usepackage{tabularx}
\usepackage{multirow}
\usepackage{bm}
\usepackage{makecell}
\usepackage{indentfirst}
\newcolumntype{C}[1]{>{\centering\let\newline\\\arraybackslash\hspace{0pt}}m{#1}}

\usepackage{color, colortbl}
\definecolor{Yellow}{rgb}{1,1,0}

\usetikzlibrary{shapes,arrows,calc,positioning}
\tikzstyle{bigblock} = [draw, fill=blue!20, rectangle, 
    minimum height=6em, minimum width=8em]
\tikzstyle{medblock} = [draw, fill=blue!20, rectangle, 
    minimum height=4em, minimum width=4em]    
\tikzstyle{mux} = [draw, fill=black!20, rectangle, 
    minimum height=5em, minimum width=0.1em]    
\tikzstyle{smallblock} = [draw, fill=blue!20, rectangle, 
    minimum height=3em, minimum width=4em]
\tikzstyle{sum} = [draw, fill=blue!20, circle, node distance=1cm]
\tikzstyle{signal} = [coordinate]
\tikzstyle{pinstyle} = [pin edge={to-,thin,black}]
\tikzstyle{block} = [draw, fill=blue!20, rectangle, 
    minimum height=3em, minimum width=6em]
\tikzstyle{blockS} = [draw, fill=blue!20, rectangle, 
    minimum height=3em, minimum width=4em]  
\tikzstyle{sum} = [draw, fill=blue!20, circle, node distance=1.5cm]
\tikzstyle{gain} = [draw, fill=blue!20, regular polygon, regular polygon sides = 3, node distance=1.25cm, shape border rotate = -90]
\tikzstyle{mult} = [draw, fill=blue!20, circle, node distance=1.25cm ,inner sep=0pt, minimum size = 0.3cm]

\tikzstyle{input} = [coordinate]
\tikzstyle{output} = [coordinate]

\newcounter{example}

\title{\LARGE Stability Analysis of
Adaptive Model Predictive Control\\
Using the Circle and Tsypkin Criteria }

\author{\large Juan A. Paredes and Dennis S. Bernstein
\thanks{Juan A. Paredes and Dennis S. Bernstein are with the Department of Aerospace Engineering, University of Michigan, Ann Arbor, MI, USA. {\tt\small \{jparedes, dsbaero\}@umich.edu}}
}

\begin{document}

\maketitle

\begin{abstract}
Absolute stability is a technique for analyzing the stability of Lur'e systems, which arise in diverse applications, such as oscillators with nonlinear damping or nonlinear stiffness.
A special class of Lur'e systems consists of self-excited systems (SES), in which bounded oscillations arise from constant inputs.
In many cases, SES can be stabilized by linear controllers, which motivates the present work, where the goal is to evaluate the effectiveness of adaptive model predictive control for Lur'e systems.
In particular, the present paper considers predictive cost adaptive control (PCAC), which is equivalent to a linear, time-variant (LTV) controller.
A closed-loop Lur'e system comprised of the positive feedback interconnection of the Lur'e system and the PCAC-based controller can thus be derived at each step.
In this work, the circle and Tsypkin criteria are used to evaluate the absolute stability of the closed-loop Lur'e system, where the adaptive controller is viewed as instantaneously linear time-invariant.
When the controller converges, the absolute stability criteria guarantee global asymptotic stability of the asymptotic closed-loop dynamics.
\end{abstract}

\section{Introduction}

Absolute stability is a technique for analyzing the stability of Lur'e systems \cite{narendra_book, leonov1996frequency}.
%
%
A Lur'e system consists of linear dynamics connected in a feedback loop with a memoryless nonlinear function.
Lur'e systems arise in diverse applications, such as oscillators with nonlinear damping or nonlinear stiffness.
More generally, viewing the nonlinear dynamics $\dot x = f(x)$ as the feedback interconnection of a vector of integrators and the multivariable memoryless nonlinearity $f,$ all nonlinear systems can be viewed as Lur'e systems.
The stability of a Lur'e system is the subject of absolute stability theory, which 
%
%
has been developed in continuous time
\cite{kalman1963,aizerman1964,yakubovich1973freq,yakubovich1973,Tomberg1989,haddad1993CT,WMHpopov,AndyPopov,arcak2002Lurie,khalil3rd,sepulchre2005,haddadbook,efimov2009,liu2010,sarkans2015,LurieHinf,contractiveLurie}
%
%
%
%
%
as well as in discrete time 
\cite{tsypkin1963, tsypkin1964,jury1964,wu1967,lee1986,haddad1994DT,WMHpopovDT1994,kapila1996,larsen2001,ahmad2011,ahmad2012,gonzaga2012,wang2014,ahmad2014,park2015,sarkans2016,park2019,Luriepreview,seiler2020,bertolin2021,bertolin2022,drummond2023,suoshea}.

A special class of Lur'e systems consists of self-excited systems (SES), which have the property that bounded oscillations arise from constant inputs
\cite{JENKINS2013167,Ding2010}.
Self-excited phenomena include flutter due to fluid-structure interaction \cite{coller,cesnik} as well as thermoacoustic oscillation in combustors \cite{awad_1986,chen_2016}.
A widely studied example of a self-excited system is the Rijke tube, in which the nonlinear heat release interacts with the linear acoustic dynamics to produce pressure oscillations \cite{rijke1859,rayleigh1878,heckl1990}.

An interesting aspect of the Rijke tube is the fact that linear controllers based on linear system identification and linearized analytical models are often effective for suppressing self-excited oscillations \cite{heckl1988,annaswamy2000,illingworth2010,epperlein2015,zalluhoglu2016, deandrade2017,RijkeTCST}.
This observation motivates the present work, where the goal is to evaluate the effectiveness of adaptive model predictive control for SES modeled by Lur'e systems.
In particular, the present paper considers predictive cost adaptive control (PCAC), which performs online closed-loop linear system identification identification;  the identified model is then used as the basis for receding-horizon optimization \cite{islamPCAC}.

In PCAC, system identification is performed during closed-loop operation by recursive least squares (RLS) \cite{islam2019recursive,mohseni2022recursive}. 
For receding-horizon optimization, quadratic programming (QP) is used in \cite{islamPCAC}.
Since state and control constraints are not crucial for SES, the present paper uses the backward propagating Riccati equation (BPRE) \cite{kwon2006receding,kwonpearson} in place of QP.
Furthermore, the resulting BPRE-based controller is equivalent to an output-feedback, model-based dynamic compensator with time-dependent gain.
The closed-loop Lur'e system comprised of the positive feedback interconnection of the Lur'e system and the PCAC-based controller can thus be derived at each step.
This observation facilitates the application of absolute stability theory for analyzing the stability of the closed-loop system.
In this work, the discrete-time circle and Tsypkin criteria \cite{tsypkin1963, tsypkin1964,sandberg1965,wu1967,lee1986,haddad1994DT,kapila1996} are used to evaluate the absolute stability of the closed-loop Lur'e system, where the adaptive controller is viewed as instantaneously linear time-invariant.
Since the closed-loop Lur'e system has time-varying linear dynamics, application of these absolute stability criteria to the instantaneous closed-loop system is heuristic.
Nevertheless, this analysis technique provides insight into the extent to which the adaptive controller stabilizes the Lur'e system.
When the controller converges, the absolute stability criteria guarantee global asymptotic stability of the asymptotic closed-loop dynamics.

The contents of the paper are as follows.
Section \ref{sec:prob} provides a statement of the control problem, which involves discrete-time, Lur'e systems.
Section \ref{sec:PCAC} describes the predictive control law considered in this paper for suppression.
Section \ref{sec:CL_DTL_PCAC} presents the closed-loop predictive controller and the closed-loop Lur'e system to apply the circle criterion to determine absolute stability.
Section \ref{sec:DTL_PCAC_exam} presents a numerical example in which the predictive controller stabilizes the oscillations of a Lur'e system and the stability of the closed-loop Lur'e system presented in Section \ref{sec:CL_DTL_PCAC} is evaluated.
Finally, Section \ref{sec:conclusion} presents conclusions. 

{\bf Notation:}
$\bfq\in\BBC$ denotes the forward-shift operator.
$x_{(i)}$ denotes the $i$th component of $x\in\BBR^n.$
${\rm spr}(G)$ denotes the maximum of the absolute values of the poles of the discrete-time transfer function $G.$
$\lambda_{\rm min} (H)$ denotes the minimum eigenvalue of Hermitian matrix $H\in\BBR^{n\times n}.$
The symmetric matrix $P\in\BBR^{n\times n}$ is positive semidefinite (resp., positive definite) if all of its eigenvalues are nonnegative (resp., positive).
$\vek X\in\BBR^{nm}$ denotes the vector formed by stacking the columns of $X\in\BBR^{n\times m}$, and $\otimes$ denotes the Kronecker product.
$I_n$ is the $n \times n$ identity matrix, and $0_{n\times m}$ is the $n\times m$ zeros matrix.
For all $r \in \BBR,$ $\rme r \isdef 10^{r}.$


\section{Statement of the Control Problem}\label{sec:prob}

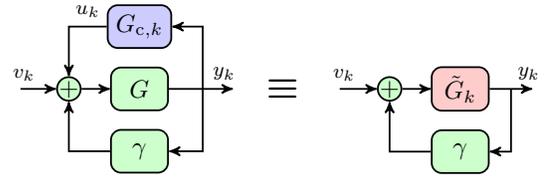
\begin{figure}[h!]
	\centering
        \vspace{0.5em}
	\resizebox{0.85\columnwidth}{!}{%
		\begin{tikzpicture}[>={stealth'}, line width = 0.25mm]
		\node [input, name=input] {};
		\node [sum, fill=green!20, right = 0.5cm of input] (sum1) {};
		\node[draw = white] at (sum1.center) {$+$};
		\node [smallblock, fill=green!20, rounded corners, right = 0.4cm of sum1, minimum height = 0.6cm, minimum width = 0.8cm] (system) {$G$};
        \node [smallblock, rounded corners, above = 0.25cm of system, minimum height = 0.6cm, minimum width = 0.8cm] (controller) {$G_{\rmc,k}$};
		\node [smallblock, fill=green!20, rounded corners, below = 0.25cm of system, minimum height = 0.6cm, minimum width = 0.8cm] (satF) {$\gamma$};
		\draw [->] (input) -- node[name=usys, xshift = -0.2cm, yshift = 0.2cm] {\footnotesize$v_k$} (sum1.west);
		\draw[->] (sum1.east) -- (system.west);
		\node [output, right = 0.9cm of system] (output) {};
		\draw [->] (system) -- node [name=y,near end, xshift = -0.2cm]{} node [near end, above, xshift = 0.1cm] {\footnotesize$y_k$}(output);
		\draw [->] (y.center) |- (satF.east);
        \draw [->] (y.center) |- (controller.east);
		\draw [->] (satF.west) -| (sum1.south);
        \draw [->] (controller.west) -| node [near start, above] {\footnotesize$u_k$} (sum1.north);
        \node [sum, fill=green!20, right = 5cm of input] (sum2) {};
        \node[draw = white] at (sum2.center) {$+$};
        \node [smallblock, fill=red!20, rounded corners, right = 0.4cm of sum2, minimum height = 0.6cm, minimum width = 0.8cm] (system2) {$\tilde{G}_k$};
        \node [smallblock, fill=green!20, rounded corners, below = 0.25cm of system2, minimum height = 0.6cm, minimum width = 0.8cm] (satF2) {$\gamma$};
		\draw [->] ([xshift = -0.7cm]sum2.center) -- node[name=usys2, xshift = -0.2cm, yshift = 0.2cm] {\footnotesize$v_k$} (sum2.west);
		\draw[->] (sum2.east) -- (system2.west);
		\node [output, right = 0.6cm of system2] (output2) {};
		\draw [->] (system2) -- node [name=y2,very near end, xshift = -0.225cm]{} node [near end, above, xshift = 0.1cm] {\footnotesize$y_k$}(output2);
		\draw [->] (y2.center) |- (satF2.east);
        \draw [->] (satF2.west) -| (sum2.south);
         \node[draw = white] at ([xshift = -1.5cm]sum2.center) {\Large$\bm \equiv$};
		\end{tikzpicture}
	}
	\caption{Closed-loop LTV control of the discrete-time Lur'e (DTL) system.  
    The LTV controller $G_{\rmc, k}$ is applied to the DTL system consisting of the linear system $G$ and feedback nonlinearity $\gamma.$
    The linear dynamics of the closed-loop Lur'e system are given by $\tilde{G}_k \isdef G (I_m -G_{\rmc,k} G)^{-1},$ which is the LTV system arising from the positive feedback interconnection of $G$ and $G_{\rmc, k}.$
    }
    \vspace{-1.75em}
	\label{fig:PC_DTL_blk_diag}
\end{figure}

We consider the control architecture shown in Figure \ref{fig:PC_DTL_blk_diag}.

Let $G(\bfq) \isdef C(\bfq I_n - A)^{-1}B$ be a
%
%
%
discrete-time, linear, time-invariant (LTI) system
with $n$th-order minimal realization $(A, B, C)$ and state $x_k \in \BBR^n$ at step $k,$ let $u_k\in\BBR^m$ be the control, let $v_k \in \BBR^m$ be the perturbation, $y_k\in\BBR^p$ be the output of $G,$ and let $\gamma\colon \BBR^p \to \BBR^m.$
Then, for all $k\ge0,$ the discrete-time Lur'e (DTL) system consisting of the linear system $G$ and feedback nonlinearity $\gamma$ shown in Figure \ref{fig:PC_DTL_blk_diag} has the closed-loop dynamics
\begin{align}
    x_{k+1} &= A x_k +  B (\gamma(y_k) + u_k + v_k), \label{xLure}\\
    y_k &= C x_k.\label{yLure}
\end{align}

At each step $k,$ the linear, time-variant (LTV) controller $G_{\rmc, k} (\bfq) \isdef C_{\rmc, k}(\bfq I_{n_\rmc}- A_{\rmc, k} )^{-1} B_{\rmc, k}$ with $n_\rmc$th-order minimal realization $(A_{\rmc, k}, B_{\rmc, k}, C_{\rmc, k})$ and state $x_{\rmc, k} \in \BBR^{n_\rmc}$ is updated by $y_k$ and generates the control $u_k$ 
such that the dynamics of the controller are given by
\begin{align}
x_{\rmc, k+1} &= \ A_{\rmc, k} x_{\rmc, k} + B_{\rmc, k} y_k, \label{eq:xCont}\\
u_k &= \ C_{\rmc, k} x_{\rmc, k}. \label{eq:yCont}
\end{align}
The objective of the controller is to provide a control signal that minimizes the output of the DTL system, that is, a control $u_k$ such that $\lim_{k \to \infty} y_k = 0.$
In this work, PCAC determines $A_{\rmc, k},B_{\rmc, k}, C_{\rmc, k},$ and $x_{\rmc, k}$ at each step $k.$

At each step $k,$ let the linear dynamics of the closed-loop Lur'e system shown in Figure \ref{fig:PC_DTL_blk_diag} be given by $\tilde{G}_k (\bfq) \isdef G(\bfq) \ [I_m -G_{\rmc, k} (\bfq) G(\bfq)]^{-1} = \tilde{C}(\bfq I_{n + n_\rmc} - \tilde{A}_k)^{-1} \tilde{B},$ 
which is a LTV system arising from the positive feedback interconnection of $G$ and $G_{\rmc, k}$ with $(n + n_\rmc)$th-order minimal realization $(\tilde{A}_k, \tilde{B}, \tilde{C}),$ state $\tilde{x}_k \isdef \matl x_k^\rmT & x_{\rmc, k}^\rmT \matr^\rmT,$ and
\begin{gather*}
\tilde{A}_k \isdef \matl A & B C_{\rmc,k} \\ B_{\rmc,k} C & A_{\rmc, k} \matr, \nn \\
\tilde{B} \isdef \matl B^\rmT & 0_{m \times n_\rmc} \matr^\rmT, \quad
\tilde{C} \isdef \matl C & 0_{p \times n_\rmc} \matr.
\end{gather*}
The dynamics of the closed-loop Lur'e system in Figure \ref{fig:PC_DTL_blk_diag} are thus given by
\begin{align}
\tilde{x}_{k+1} &= \ \tilde{A}_k \tilde{x}_k + \tilde{B} (\gamma(y_k) + v_k), \label{eq:xLureEq}\\
y_k &= \ \tilde{C} \tilde{x}_k. \label{eq:yLureEq}
\end{align}

\vspace{-.2in}

\section{Predictive Cost Adaptive Control} \label{sec:PCAC}

PCAC is briefly reviewed in this section.
Subsection \ref{subsec:ID} describes the technique used for online identification, namely, RLS with variable-rate forgetting based on the F-test \cite{mohseni2022recursive}.
Subsection \ref{subsec:bocf} presents the block observable canonical form (BOCF), which is used to represent the input-output dynamics model as a state space model whose state is given explicitly in terms of inputs, outputs, and model-coefficient estimates.
Subsection \ref{subsec:bpre} reviews the backward-propagating Riccai equation (BPRE) technique for receding-horizon optimization.
Using BOCF, the full-state feedback controller obtained using BPRE is implementable as an output-feedback dynamic compensator.

\subsection{Online Identification Using Recursive Least Squares with Variable-Rate Forgetting Based on the F-Test} \label{subsec:ID}

Let $\hat n\ge 0$ and, for all $k\ge 0,$ let $F_{\rmm,1,k},\hdots, F_{\rmm,\hat n,k}\in\BBR^{p\times p}$ and $G_{\rmm,1,k},\hdots, G_{\rmm,\hat n,k}\in\BBR^{p\times m}$ be the coefficient matrices to be estimated using RLS.
Furthermore, let $\hat y_k\in\BBR^p$ be an estimate of $y_k$ defined  by
\begin{equation}
\hat y_k\isdef -\sum_{i=1}^{\hat n}  F_{\rmm,i,k}   y_{k-i} + \sum_{i=1}^{\hat n} {G}_{\rmm,i,k} u_{k-i},
\label{eq:yhat}
\end{equation}
where  
\begin{gather}
   y_{-\hat n}=\cdots= y_{-1}=0,\\ u_{-\hat n}=\cdots=u_{-1}=0. 
\end{gather}
Using the identity ${\rm vec} (XY) = (Y^\rmT \otimes I) {\rm vec} X,$ it follows from  \eqref{eq:yhat} that, for all $k\ge 0,$ 
\begin{equation}
    \hat y_k = \phi_k \theta_k,
    \label{eq:yhat_phi}
\end{equation}
where
\begin{align}
     \theta_k \isdef & \ \matl \theta_{F_\rmm, k}^\rmT & \theta_{G_\rmm, k}^\rmT \matr^\rmT \in\BBR^{\hat np(m+p)},\\
     \theta_{F_\rmm, k} \isdef & \ {\rm vec}\matl  F_{\rmm,1,k}&\cdots& F_{\rmm,\hat n,k} \matr \in\BBR^{\hat n p^2}, \\
     \theta_{G_\rmm, k} \isdef & \ {\rm vec}\matl  G_{\rmm,1,k}&\cdots& G_{\rmm,\hat n,k} \matr \in\BBR^{\hat n pm}, \\
     \phi_k \isdef & \matl -  y_{k-1}^\rmT&\cdots&- y_{k-\hat n}^\rmT&u_{k-1}^\rmT&\cdots& u_{k-\hat n}^\rmT\matr\otimes I_p \nn \\ &\in\BBR^{p\times \hat np(m+p)}.
\label{eq:phi_kkka}
 \end{align}

To determine the update equations for $\theta_k$, for all $k\ge 0$, define $e_k\colon\BBR^{\hat np(m+p)}\to\BBR^p$ by
\begin{equation}
    e_k(\bar \theta) \isdef y_k - \phi_k \bar \theta,
    \label{eq:ekkea}
\end{equation}
where $\bar \theta\in\BBR^{\hat np(m+p)}.$ 
Using  \eqref{eq:yhat_phi}, the \textit{identification error} at step $k$ is defined by
\begin{equation}
 e_k(\theta_k)= y_k-\hat y_k.  
\end{equation}
For all $k\ge 0$, the RLS cumulative cost $J_k\colon\BBR^{\hat np(m+p)}\to[0,\infty)$ is defined by \cite{islam2019recursive}
\begin{equation}
J_k(\bar \theta) \isdef \sum_{i=0}^k \frac{\rho_i}{\rho_k} e_i^\rmT(\bar \theta) e_i(\bar \theta) + \frac{1}{\rho_k} (\bar\theta -\theta_0)^\rmT \Psi_0^{-1}(\bar\theta-\theta_0),
\label{Jkdefn}
\end{equation}
where $\Psi_0\in\BBR^{\hat np(m+p)\times \hat np(m+p)}$ is  positive definite, $\theta_0\in\BBR^{\hat n p(m+p)}$ is the initial estimate of the coefficient vector, and, for all $i\ge 0,$
\begin{equation}
  \rho_i \isdef \prod_{j=0}^i \lambda_j^{-1}.  
\end{equation}
For all $j\ge 0$, the parameter $\lambda_j\in(0,1]$ is the forgetting factor defined by $\lambda_j\isdef\beta_j^{-1}$, where
\begin{equation}
 \beta_j \isdef \begin{cases}
     1, & j<\tau_\rmd,\\
     1 + \eta \bar{\beta}_j,& j\ge \tau_\rmd,
 \end{cases}    
\end{equation}
\footnotesize
\begin{equation}
    \bar{\beta}_j \isdef g(e_{j-\tau_\rmd}(\theta_{j-\tau_\rmd}),\hdots,e_j(\theta_j)) \textbf{1}\big(g(e_{j-\tau_\rmd}(\theta_{j-\tau_\rmd}),\hdots,e_j(\theta_j))\big),
\end{equation}
\normalsize
and  $\tau_\rmd> p$, $\eta>0$,  $\textbf{1}\colon \BBR\to\{0,1\}$ is the unit step function, and $g$ is the function of past RLS identification errors given by (10) in \cite{mohseni2022recursive} for $p = 1$ and (13) in \cite{mohseni2022recursive} for $p > 1.$
Note that $g$ includes forgetting terms based on the inverse cumulative distribution function of the F-distribution and depends on $\tau_\rmd,$ $\tau_\rmn\in[p,\tau_\rmd),$ and \textit{significance level} $\alpha\in (0,1].$

Finally, for all $k\ge0$, the unique global minimizer of $J_k$ is given by \cite{islam2019recursive}
\begin{equation}
    \theta_{k+1} = \theta_k +\Psi_{k+1} \phi_k^\rmT (y_k - \phi_k \theta_k),
\end{equation}
where 
\begin{align}
  \Psi_{k+1} &\isdef  \beta_k \Psi_k - \beta_k \Psi_k \phi_k^\rmT (\tfrac{1}{\beta_k}I_p + \phi_k  \Psi_k \phi_k^\rmT)^{-1} \phi_k  \Psi_k,
\end{align}
and $\Psi_0$ is the performance-regularization weighting in \eqref{Jkdefn}.
Additional details concerning RLS with forgetting based on the F-distribution are given in \cite{mohseni2022recursive}.


\subsection{Input-Output Model and the Block Observable Canonical Form} \label{subsec:bocf}

Considering the estimate $\hat y_k$ of $y_k$ given by \eqref{eq:yhat}, it follows that, for all $k\ge0,$
\begin{equation}
y_{k} \approx -\sum_{i=1}^{\hat n}  F_{\rmm,i,k}   y_{k-i} + \sum_{i=1}^{\hat n} {G}_{\rmm,i,k} u_{k-i}.
\label{eq:ykapp}
\end{equation}
Viewing \eqref{eq:ykapp} as an equality, it follows that, for all $k\ge 0,$ the block observable canonical form (BOCF) state-space realization of \eqref{eq:ykapp} is given by  \cite{polderman1989state}
\begin{align}
 x_{\rmm,k+1} &=   A_{\rmm,k}  x_{\rmm,k} +  B_{\rmm,k} u_k,\label{eq:xmssAB}\\
 y_{k} &=  C_\rmm   x_{\rmm, k},
\label{eq:yhatCxm}
\end{align}
where
\begin{gather}
 A_{\rmm,k} \isdef \matl - F_{\rmm,1,k+1} & I_p & \cdots & \cdots & 0_{p\times p}\\
- F_{\rmm,2,k+1} & 0_{p\times p} & \ddots & & \vdots\\
\vdots & {\vdots} & \ddots & \ddots & 0_{p\times p} \\
\vdots & \vdots &  & \ddots & I_p\\
- F_{\rmm,\hat n,k+1} & 0_{p\times p} & \cdots &\cdots & 0_{p\times p}
\matr\in\BBR^{\hat np\times \hat n p},\label{eq:ABBA_1}\\
B_{\rmm,k}\isdef \matl  G_{\rmm,1,k+1} \\
 G_{\rmm,2,k+1}\\
\vdots\\
 G_{\rmm,\hat n,k+1}
\matr \in\BBR^{\hat n p \times m},\label{eq:ABBA_2}
\end{gather}
\vspace{-1.5em}
\begin{gather}
  C_\rmm\isdef \matl I_p & 0_{p\times p} & \cdots & 0_{p\times p} \matr\in\BBR^{p\times \hat n p},
  \label{eq:cmmx}
\end{gather}
and 
\begin{equation}
    x_{\rmm,k} \isdef \matl x_{\rmm,k(1)}\\\vdots\\ x_{\rmm,k(\hat n)}\matr \in\BBR^{\hat n p},
    \label{eq:xmkkx}
\end{equation}
where
$x_{\rmm,k(1)} \isdef  y_{k},$
and, for all $j=2,\ldots,\hat n,$
\begin{align}
 x_{\rmm,k(j)} \isdef & -\sum_{i=1}^{\hat n -j +1}  F_{\rmm,i+j-1,k+1}  y_{k-i} \nn \\
 &+ \sum_{i=1}^{\hat n -j+1}  G_{\rmm,i+j-1,k+1} u_{k-i}.
 \label{eq:xnkk}
\end{align}
Note that multiplying both sides of \eqref{eq:xmssAB} by $C_\rmm$ and using \eqref{eq:yhatCxm}--\eqref{eq:xnkk} implies that, for all $k\ge 0,$
\small
\begin{align}
     y_{k+1}&=C_\rmm x_{\rmm,k+1}\nn\\
     &= C_\rmm(A_{\rmm,k}  x_{\rmm,k} +  B_{\rmm,k} u_k)\nn\\
    &=-F_{\rmm,1,k+1}  x_{\rmm,k(1)} + x_{\rmm,k(2)} +G_{\rmm,1,k+1} u_k \nn\\
    &=-F_{\rmm,1,k+1}  y_{k}  -\sum_{i=1}^{\hat n -1}  F_{\rmm,i+1,k+1}  y_{k-i} \nn\\
    &+ \sum_{i=1}^{\hat n -1}  G_{\rmm,i+1,k+1} u_{k-i}+G_{\rmm,1,k+1} u_k \nn\\
    &=  -\sum_{i=1}^{\hat n }  F_{\rmm,i,k+1}  y_{k+1-i} + \sum_{i=1}^{\hat n }  G_{\rmm,i,k+1} u_{k+1-i},
\end{align}
\normalsize
which is approximately equivalent to \eqref{eq:ykapp} with $k$ in \eqref{eq:ykapp} replaced by $k+1$.


\subsection{Receding-Horizon Control with Backward-Propagating Riccati Equation (BPRE)} \label{subsec:bpre}

In this section, we use receding-horizon optimization to determine the requested control $u_{k+1}$ and thus the implemented control $u_{k+1}$, as discussed in Section \ref{sec:prob}.
Let $\ell\ge 1$ be the horizon, and, for all $k\ge0$ and all $j=1,\ldots,\ell,$ consider the state-space prediction model
\begin{equation}
    x_{\rmm,k|j+1} =   A_{\rmm,k}  x_{\rmm,k|j} +  B_{\rmm,k}  u_{k|j},
\end{equation}
where $A_{\rmm,k}$ and $B_{\rmm,k}$ are given by \eqref{eq:ABBA_1} and \eqref{eq:ABBA_2}, respectively,
$x_{\rmm,k|j}\in\BBR^{\hat n p}$ is the $j$-step predicted state, $u_{k|j}\in\BBR^m$ is the $j$-step predicted control, and the initial conditions are
\begin{equation}
 x_{\rmm,k|1} \isdef  x_{\rmm,k+1},\quad u_{k|1}\isdef  u_{k+1}. \label{eq:xmtuk1} 
\end{equation}
Note that, at each step  $k\ge0$, after obtaining the measurement $y_k$, $x_{\rmm,k+1}$ is computed using \eqref{eq:xmssAB}, where $u_k$ is the implemented control at step $k.$
Furthermore,  $u_{k+1}$,
which is determined below, is the requested control at step $k+1$.
For all $k\ge0$, define the performance index
\begin{align}
\SJ_k(&u_{k|1},\hdots, u_{k|\ell}) \nn \\ 
\isdef & \ \half \sum_{j=1}^{\ell} ( x_{\rmm,k|j}^\rmT R_{1,k|j}  x_{\rmm,k|j} +  u_{k|j}^\rmT R_{2,k|j}   u_{k|j}) \nn \\
&+ \half  x_{\rmm,k|\ell+1}^\rmT P_{k|\ell+1} x_{\rmm,k|\ell+1},\label{SJk}
\end{align}
where the terminal weighting $P_{k|\ell+1}\in\BBR^{\hat np\times \hat np}$ is positive semidefinite and,
for all  $j=1,\ldots,\ell,$  
$R_{1,k|j}\in\BBR^{\hat np\times \hat n p}$ is the positive semidefinite state weighting and $R_{2,k|j}\in\BBR^{m\times m}$ is the positive definite control weighting.
The first term in \eqref{SJk} can be written as
\begin{equation}
    x_{\rmm,k|j}^\rmT R_{1,k|j}  x_{\rmm,k|j} = z_k^\rmT z_k,
\end{equation}
where $z_k\in\BBR^p$ is defined by
\begin{equation}
    z_k \isdef E_{1,k|j}x_{\rmm,k|j},
\end{equation}
and $E_{1,k|j}\in\BBR^{p\times \hat n p}$ is defined such that
\begin{equation}
    R_{1,k|j} = E_{1,k|j}^\rmT E_{1,k|j}.
\end{equation}
With this notation, $z_k$ is the performance variable.


For all $k\ge0$ and  $j=\ell,\ell-1,\ldots,2,$ let $P_{k|j}$ be given by
\begin{align}
P_{k|j} &= \ A_{\rmm,k}^\rmT P_{k|j+1}   \left(A_{\rmm,k} -  B_{\rmm,k} \Gamma_{k|j} \right) + R_{1,k|j}, \\
%
\Gamma_{k|j} \isdef & \ (R_{2,k|j} +   B_{\rmm,k}^\rmT P_{k|j+1}  B_{\rmm,k})^{-1}   B_{\rmm,k}^\rmT P_{k|j+1}   A_{\rmm,k}.\nn
\end{align}
Then, for  all $k\ge0$ and all $j=1,\ldots,\ell,$ the requested optimal  control is given by
\begin{equation}
     u_{k|j} = K_{k|j} x_{\rmm,k|j},
     \label{eq:tuKxm}
\end{equation}
%
%
\begin{equation}
    K_{k|j} \isdef -(R_{2,k|j} +   B_{\rmm,k}^\rmT P_{k|j+1}   B_{\rmm,k})^{-1}  B_{\rmm,k}^\rmT P_{k|j+1}  A_{\rmm,k}.\nn
    \label{eq:Kkjj}
\end{equation}
For all $k\ge0,$ define $K_{k+1}\isdef K_{k|1}$.
Then, for all $k\ge0,$  \eqref{eq:xmtuk1} and \eqref{eq:tuKxm} imply that 
\begin{align}
  u_{k+1}  &=  K_{k+1} x_{\rmm,k+1}, 
  \label{eq:ukxkkx}
\end{align}
which, combined with \eqref{eq:Kkjj}, implies that
\small
\begin{equation}
u_{k+1}=  -(R_{2,k|1} +   B_{\rmm,k}^\rmT P_{k|2}  B_{\rmm,k})^{-1}  B_{\rmm,k}^\rmT P_{k|2}   A_{\rmm,k} x_{\rmm,k+1}.
\end{equation}
\normalsize
%
%

Note that the initial control $u_0\in\BBR^m$ is not computed and must be specified. 
In this work, $u_0 = 0.$
In addition, note that, for all $k\ge0,$ the requested control $u_{k+1}$ is computed during the interval $[kT_\rms,(k+1)T_\rms)$ using the measurement $y_k$ and the implemented control $u_k.$
Furthermore, note that, for all $k\ge0$ and all $j = 2,\ldots,\ell,$ $K_{k|j}$, $x_{\rmm,k|j},$ and $u_{k|j}$ need not be computed, in accordance with receding-horizon control. 
Finally, for all examples in this paper, we choose $R_{1,k|j}$, $R_{2,k|j}$, $E_{1,k|j}$ and $P_{k|\ell+1}$  to be independent of $k$ and $j,$ and we thus write $R_1$, $R_2$, $E_1$, and $P_{\ell+1},$ respectively.


\section{Closed-loop Lur'e system for absolute stability analysis} \label{sec:CL_DTL_PCAC}

To derive the dynamics of the closed-loop Lur'e system, note that
\eqref{eq:xmssAB} and \eqref{eq:yhatCxm} imply that
\begin{equation}\label{eq:CL_C_1}
    x_{\rmm, k+1} =  (A_{\rmm, k} - F_k C_\rmm) x_{\rmm, k} + B_{\rmm, k} u_k + F_k y_k, 
\end{equation}
where
\begin{equation*}
F_k \isdef \matl -F_{\rmm, 1, k+1}^\rmT & \cdots & -F_{\rmm, \hat{n}, k+1}^\rmT \matr^\rmT \in \BBR^{\hat{n} p \times p}.
\end{equation*}
Then, it follows from \eqref{eq:ukxkkx} and \eqref{eq:CL_C_1}
\begin{equation}\label{eq:CL_C_2}
    x_{\rmm, k+1} =  (A_{\rmm, k} - F_k C_\rmm + B_{\rmm, k} K_k) x_{\rmm, k} + F_k y_k. 
\end{equation}
%
%
Hence, replacing $n_\rmc,$ $x_{\rmc, k},$ \eqref{eq:xCont} and \eqref{eq:yCont} with $\hat{n} p,$ $x_{\rmm,k},$ \eqref{eq:ukxkkx} and \eqref{eq:CL_C_2}, respectively, yields the closed-loop dynamics of the PCAC-based controller
\begin{align}
x_{\rmc, k+1} = x_{\rmm, k+1} &= \ A_{\rmc, k} x_{\rmm, k} + B_{\rmc, k} y_k,\label{eq:xPCAC}\\
u_k &= \ C_{\rmc, k} x_{\rmm, k},\label{eq:yPCAC}
\end{align}
where
\begin{equation*}
A_{\rmc, k} \isdef A_{\rmm, k} - F_k C_\rmm + B_{\rmm, k} K_k, \
B_{\rmc, k} \isdef F_k, \
C_{\rmc, k} \isdef K_k.
\end{equation*}

Next, we analyze the stability of the closed-loop Lur'e system shown in Figure \ref{fig:PC_DTL_blk_diag} with the controller \eqref{eq:xPCAC}, \eqref{eq:yPCAC}.
Since the feedback controller is LTV, absolute stability tests such as the circle and Tsypkin criteria, which assume that the linear dynamics are LTI, are not applicable.
However, as a heuristic criterion, we apply these criteria instantaneously, which provides insight into the ability of PCAC to  stabilize the Lur'e system.
When the controller converges, the absolute stability criteria guarantee global asymptotic stability of the asymptotic closed-loop dynamics.
The absolute stability criteria used in this work are summarized in the Appendix;
in particular, Theorem \ref{theo:AbsStabCC} is the discrete-time circle criterion, and Theorem \ref{theo:AbsStabTC} is the Tsypkin criterion.
Furthermore, sector-bounded (SB) and diagonal,  increasing, and sector-bounded (DISB) nonlinearities are defined by Definitions \ref{def_sector_bounded} and \ref{def_mono_sector_bounded}, respectively.

To apply the circle criterion, 
let $\gamma$ be SB with sector bound [$M_1, M_2$], and, for all $k\ge0,$ define
\begin{align*}
H_k \isdef & \ [I_m - M_2 \tilde{G}_k] [I_m - M_1 \tilde{G}_k]^{-1},\\
\alpha_{{\rm CC}, k} \isdef & \ {\rm spr} (H_k),\\
\beta_{{\rm CC}, k} \isdef & \min_{\psi \in [0, \pi]} \left(\lambda_{\rm min} \left[ H_k (e^{\jmath \psi}) + H_k^\rmT (e^{-\jmath \psi}) \right] \right).
\end{align*}
At step $k,$ 
if $\alpha_{{\rm CC}, k} < 1,$ then (CC1) of Theorem \ref{theo:AbsStabCC} is instantaneously satisfied by the closed-loop Lur'e system,
and, if $\beta_{{\rm CC}, k} > 0,$ then (CC2) of Theorem \ref{theo:AbsStabCC} is instantaneously satisfied by the closed-loop Lur'e system.
Hence, Theorem \ref{theo:AbsStabCC} implies that, if  $\alpha_{{\rm CC}, k} < 1$ and $\beta_{{\rm CC}, k} > 0,$ then the closed-loop Lur'e system shown in Figure \ref{fig:PC_DTL_blk_diag} with the PCAC controller   satisfies (CC1) and (CC2)  at step $k.$

To apply the Tsypkin criterion,
let $m = p,$~let $\gamma$
 be DISB with sector bound [$0, M$], let $N \isdef {\rm diag} (N_1, N_2, \ldots, N_m)$ be positive definite, and, for all $k\ge0,$ define
\begin{align*}
L_{N, k}(\bfq) \isdef & \ M^{-1} - [I_m + (1 - \bfq^{-1}) N] \ \tilde{G}_k(\bfq),\\
\zeta_{1,k} \isdef & \ {\rm det} (\tilde{C} \tilde{A}_k^{-1} \tilde{B}),\\
\zeta_{2,N,k} \isdef & \ {\rm rank} ({\rm obsv} ((\tilde{A}_k, \tilde{C} + N \tilde{C} - N \tilde{C} \tilde{A}_k^{-1}))),\\
\zeta_{3,N,k} \isdef & \lim_{\bfz \to \infty} [L_N (\bfz) + L_N^\rmT (\bfz)],\\
\alpha_{{\rm TC}, N, k} \isdef & \ {\rm spr} (L_{N, k}),\\
%
\beta_{{\rm TC}, N, k} \isdef & \min_{\psi \in [0, \pi]} \left( \lambda_{\rm min} \left[ L_{N, k} (e^{\jmath \psi}) + L_{N, k}^\rmT (e^{-\jmath \psi}) \right]\right).
\end{align*}
At step $k,$
if $\zeta_{1,k} \neq 0,$ $\zeta_{2,N,k} = n + \hat{n}p,$ and $\zeta_{3,N,k} > 0,$ then (TC1) of Theorem \ref{theo:AbsStabTC} is instantaneously satisfied by the closed-loop Lur'e system,
if $\alpha_{{\rm TC}, N, k} < 1,$ then (TC2) of Theorem \ref{theo:AbsStabTC} is instantaneously satisfied by the closed-loop Lur'e system,
and, if $\beta_{{\rm TC}, N, k} > 0,$ then (TC3) of Theorem \ref{theo:AbsStabTC} is instantaneously satisfied by the closed-loop Lur'e system.
Hence, Theorem \ref{theo:AbsStabTC} implies that, if  $\zeta_{1,k} \neq 0,$ $\zeta_{2,N,k} = n + \hat{n}p,$ $\zeta_{3,N,k} > 0,$ $\alpha_{{\rm TC}, N, k} < 1$ and $\beta_{{\rm TC}, N, k} > 0,$ then the closed-loop Lur'e system shown in Figure \ref{fig:PC_DTL_blk_diag} with the PCAC controller instantaneously satisfies (TC1), (TC2), and (TC3)  at step $k.$

These conditions are evaluated in Section \ref{sec:DTL_PCAC_exam} to determine whether PCAC yields a globally asymptotically stable (GAS) closed-loop Lur'e system.


\section{Numerical Example} \label{sec:DTL_PCAC_exam}

In this section, PCAC is applied to a DTL system, and the resulting closed-loop system is analyzed by using absolute stability criteria.
In particular, the circle criterion given by Theorem \ref{theo:AbsStabCC} and the Tsypkin criterion given by Theorem \ref{theo:AbsStabTC} are used to evaluate the stability of the closed-loop system at each step.
Since the closed-loop Lur'e system has time-varying linear dynamics, application of these absolute stability criteria is heuristic.
Nevertheless, this analysis technique provides insight into the ability of the adaptive controller to stabilize the Lur'e system.
When the controller converges, the absolute stability criteria guarantee global asymptotic stability of the asymptotic closed-loop dynamics.

Example \ref{ex_1} features a bounded, monotonic nonlinearity that lies in the first and third quadrants.
This system exhibits self-oscillation.
Hence, the simulations begin in open-loop operation to allow the self-excited oscillations to fully develop before PCAC is applied.  
In this example, the RLS hyperparameters are  
\begin{gather*}
  \hat{n} = 10, \quad \theta_0 = 10^{-10}\, 1_{2 \hat{n}\times 1},\quad \Psi_0 = 10^{-4} I_{2 \hat{n}}, \\ \tau_\rmn = 40,\quad \tau_\rmd = 200, \quad \eta = 0.1,\quad \alpha = 0.001,
\end{gather*}
and the BPRE hyperparameters are
\begin{gather*}
    \ell = 20, \quad P_{\ell+1}= \diag(1, 0_{1\times \hat{n}-1}), \\
    R_1 =  \diag(1,0_{1\times \hat{n}-1}), \quad R_2= 10^{-4}.
\end{gather*}
The exogenous input $v_k$ is used to enhance persistency and thus enhance the ability of the closed-loop Lur'e system to satisfy the absolute stability criteria. 
Three cases are considered, namely, for all $k \ge 0,$ $v_k = 0,$ $v_k = v_{{\rm imp}, k},$ and $v_k = v_{{\rm rand}, k},$ where
%
%
\begin{align} 
    v_{{\rm imp}, k} \isdef & \begin{cases}
      1 &  k \in \{1000, 1400, 1800\} \\
      -1 &  k \in \{1200, 1600, 2000\} \\
      0 & \mbox{otherwise},
    \end{cases} \label{eq:v_imp_ex_1} \\
    v_{{\rm rand}, k} \isdef & \begin{cases}
      \sigma_k &  k \in [1000, 1500] \\
      0 & \mbox{otherwise},
    \end{cases} \label{eq:v_randn_ex_1}
\end{align}
and $\sigma_k$ is a Gaussian random variable with mean 0 and standard deviation 1.

\begin{figure}[h!]
\vspace{-0.5em}
    \centering
    \resizebox{0.85\columnwidth}{!}{%
    \includegraphics{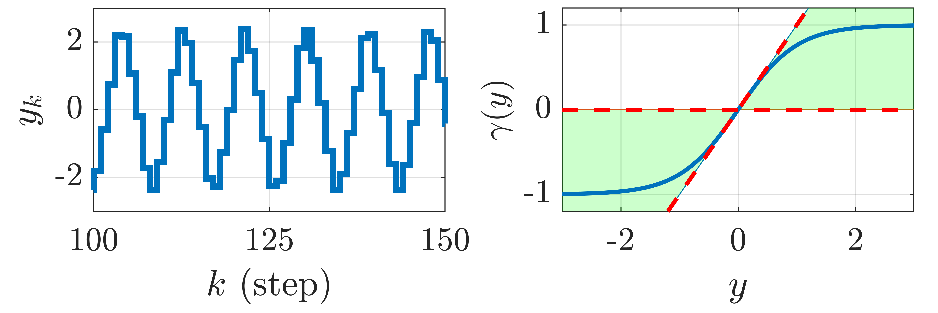}
    }
    \vspace{-0.25em}
    \caption{Example \ref{ex_1}: Open-loop response of the DTL system in Example \ref{ex_1} for $k\in[100, 150]$.  $\gamma$ is  SB and DISB with sector bound $[0, 1],$ as indicated by the red, dashed line segments and the green-shaded region.
    %
    %
    }
    \label{fig:ex1_OL}
    \vspace{-2em}
\end{figure}
\begin{figure}[h!]
    \centering
    \resizebox{\columnwidth}{!}{%
    \includegraphics{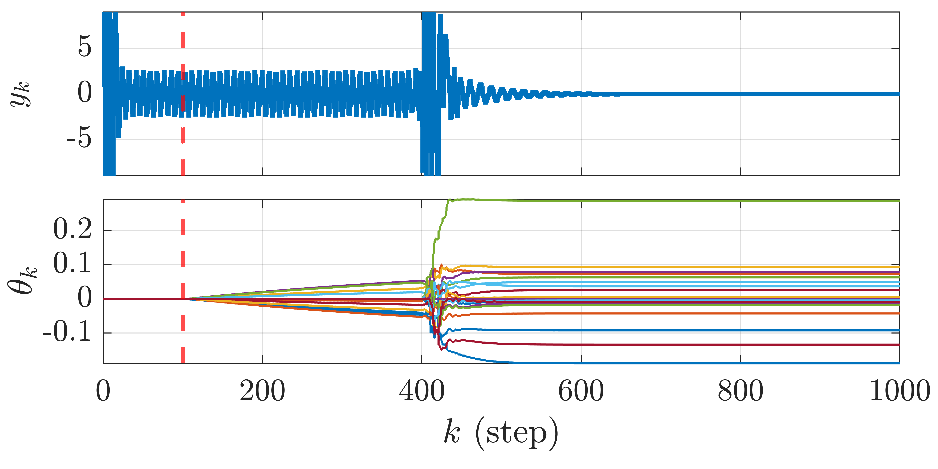}
    }
    \vspace{-2em}
    \caption{Example \ref{ex_1}: 
    DTL system output $y$ and the adaptive controller coefficients $\theta_k$ for $x_0 = 1000B$ and $v_k = 0$.  
    The simulation transitions from open-loop operation to closed-loop operation at the step indicated by the vertical, dashed red line.
    }
    \label{fig:ex1_time} 
    \vspace{-1.75em}
\end{figure}
\begin{figure}[h!]
    \centering
    \resizebox{\columnwidth}{!}{%
    \includegraphics{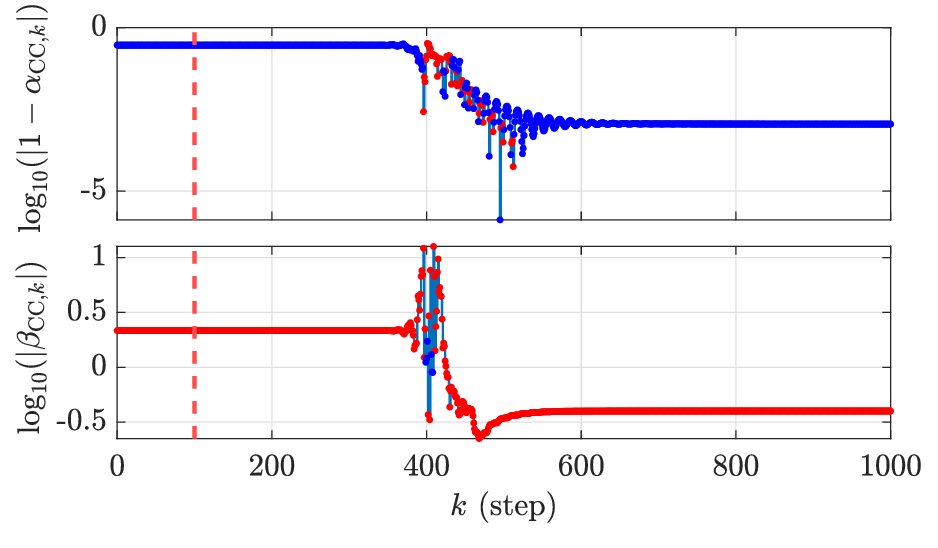}
    }
    \vspace{-2em}
    \caption{Example \ref{ex_1}:
    Evaluation of (CC1) and (CC2) for $x_0 = 1000B$ and $v_k = 0.$  
    Values in red correspond to steps at which $1 - \alpha_{{\rm CC}, k}$ and  $\beta_{{\rm CC}, k}$ are nonpositive.
    Values in blue correspond to steps at which $1 - \alpha_{{\rm CC}, k}$ and  $\beta_{{\rm CC}, k}$ are positive.
    The closed-loop system satisfies (CC1) and (CC2) when the values in both plots are green.
    For this case, the circle criterion is not satisfied at most steps.
    }
    \label{fig:ex1_circ}
    \vspace{-1em}
\end{figure}
\begin{figure}[h!]
    \centering
    \vspace{0.5em}
    \resizebox{\columnwidth}{!}{%
    \includegraphics{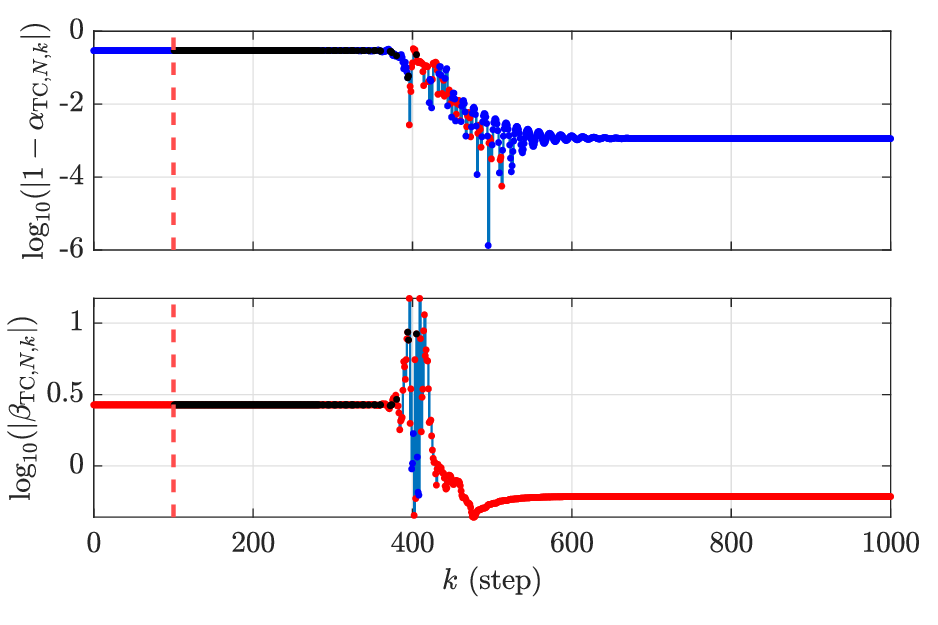}
    }
    \vspace{-2.5em}
    \caption{Example \ref{ex_1}:
    Evaluation of the Tsypkin criterion  for $x_0 = 1000B,$ and $v_k = 0$ with $N = 0.08.$ 
    Values not in black correspond to steps at which $\zeta_{1,k} \neq 0,$ $\zeta_{2,N,k} = n + \hat{n}p,$ and $\zeta_{3,N,k} > 0.$
    Values in red correspond to steps at which $1 - \alpha_{{\rm TC}, N, k}$ and $\beta_{{\rm TC}, N, k}$ are nonpositive.
    Values in blue correspond to steps at which $1 - \alpha_{{\rm TC}, N, k}$ and $\beta_{{\rm TC}, N, k}$ are positive.
    The closed-loop system satisfies (TC1), (TC2) and (TC3) when the values in both plots are green.
    For this case, the Tsypkin criterion is not satisfied at most steps.
    }
    \label{fig:ex1_tsyp}
    \vspace{-1.25em}
\end{figure}
\begin{figure}[h!]
    \centering
    \resizebox{0.9\columnwidth}{!}{%
    \includegraphics{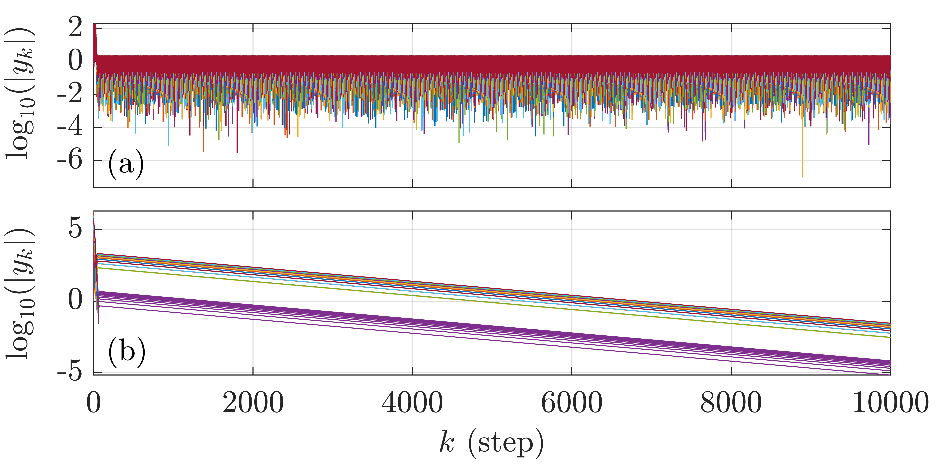}
    }
    \vspace{-0.5em}
    \caption{Example \ref{ex_1}:
    (a) shows the output $y$ of the open-loop DTL system with linear dynamics $G$.
    (b) shows the output $y$ of the closed-loop DTL system with linear dynamics $\tilde{G}_{1000}$.
    %
    %
    Both plots show the response for all $x_0 \in \{-\rme 6, - 9 \rme 5, \ldots, 9 \rme 5, \rme 6 \} \times\{-\rme 6, - 9 \rme 5, \ldots, 9 \rme 5, \rme 6 \}.$   
    }
    \label{fig:ex1_roa}
    \vspace{-1.5em}
\end{figure}
\begin{figure}[h!]
    \centering
    \vspace{0.5em}
    \resizebox{\columnwidth}{!}{%
    \includegraphics{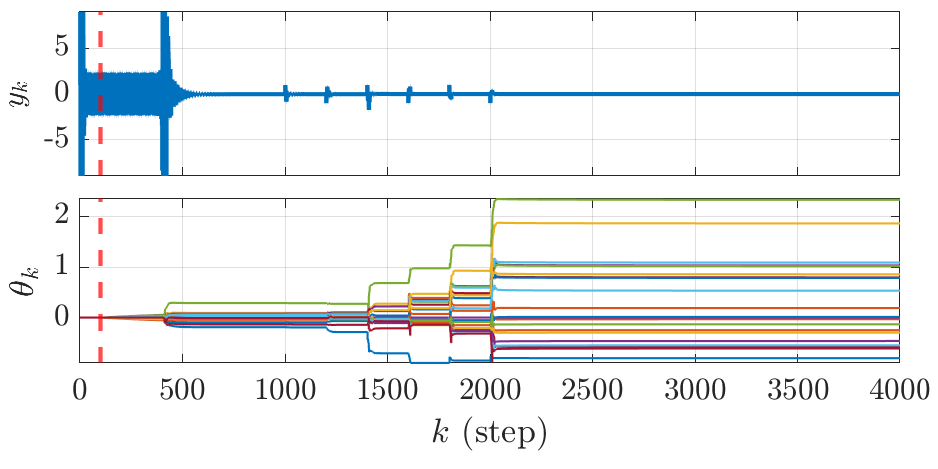}
    }
    \vspace{-2em}
    \caption{Example \ref{ex_1}: 
    DTL system output $y$ and the adaptive controller coefficients $\theta_k$ for $x_0 = 1000B$ and $v_k = v_{{\rm imp}, k}$.
    %
    The simulation transitions from open-loop operation to closed-loop operation at the step indicated by the vertical, dashed red line.
    }
    \label{fig:ex1_time_Vimp}
    \vspace{-0.5em}
\end{figure}
\begin{figure}[h!]
    \centering
    \vspace{0.5em}
    \resizebox{\columnwidth}{!}{
    \includegraphics{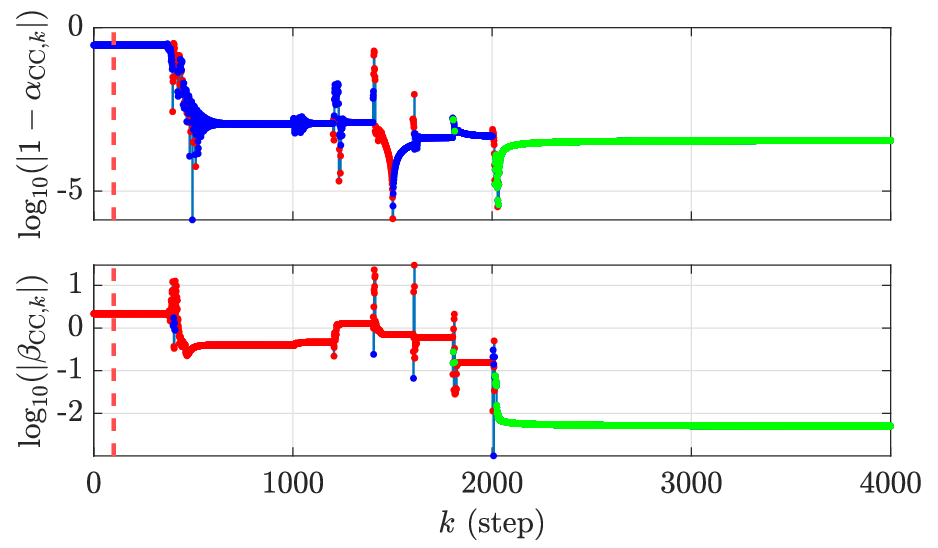}
    }
    \vspace{-2em}
    \caption{Example \ref{ex_1}:
    Evaluation of (CC1) and (CC2) for $x_0 = 1000B$ and $v_k = v_{{\rm imp}, k}$.  
    Values in red correspond to steps at which $1 - \alpha_{{\rm CC}, k}$ and  $\beta_{{\rm CC}, k}$ are nonpositive.
    Values in blue correspond to steps at which $1 - \alpha_{{\rm CC}, k}$ and  $\beta_{{\rm CC}, k}$ are positive.
    The closed-loop system satisfies (CC1) and (CC2) when the values in both plots are green.
    For this case, the circle criterion is satisfied for all steps  $k\ge2000.$ 
    }
    \label{fig:ex1_circ_Vimp}
    \vspace{-1.5 em}
\end{figure}
\begin{figure}[h!]
    \centering
    \resizebox{\columnwidth}{!}{%
    \includegraphics{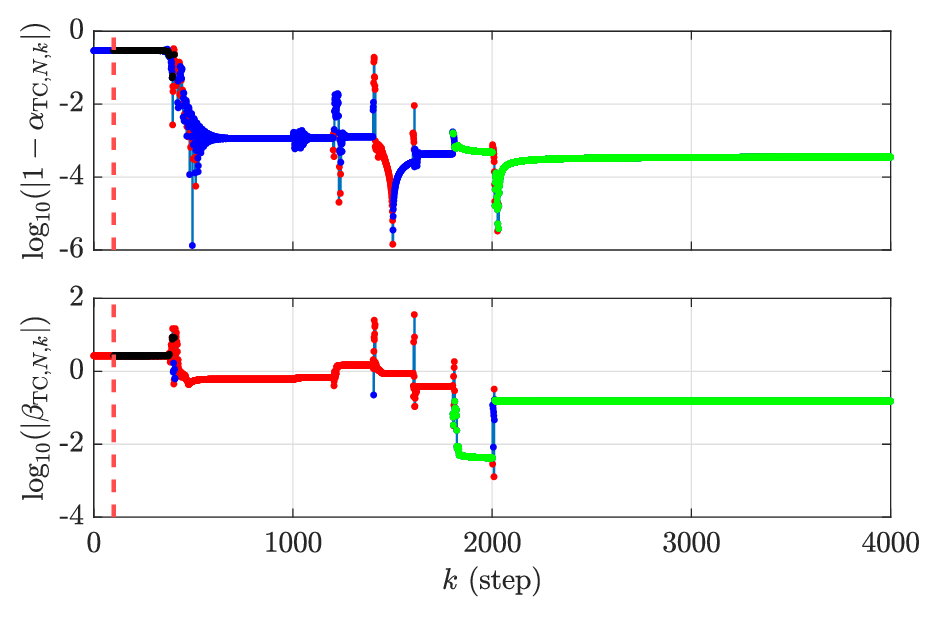}
    }
    \vspace{-2.5em}
    \caption{Example \ref{ex_1}:
    Evaluation of the Tsypkin criterion  for $x_0 = 1000B$ and $v_k = v_{{\rm imp}, k}$ with $N = 0.08.$ 
    Values not in black correspond to steps at which $\zeta_{1,k} \neq 0,$ $\zeta_{2,N,k} = n + \hat{n}p,$ and $\zeta_{3,N,k} > 0.$
    Values in red correspond to steps at which $1 - \alpha_{{\rm TC}, N, k}$ and $\beta_{{\rm TC}, N, k}$ are nonpositive.
    Values in blue correspond to steps at which $1 - \alpha_{{\rm TC}, N, k}$ and $\beta_{{\rm TC}, N, k}$ are positive.
    The closed-loop system satisfies (TC1), (TC2) and (TC3) when the values in both plots are green.
    For this case, the Tsypkin criterion is satisfied for all $k\ge2000$.
    }
    \label{fig:ex1_tsyp_Vimp}
    \vspace{-1.5em}
\end{figure}
\begin{figure}[h!]
    \centering
    \resizebox{\columnwidth}{!}{%
    \includegraphics{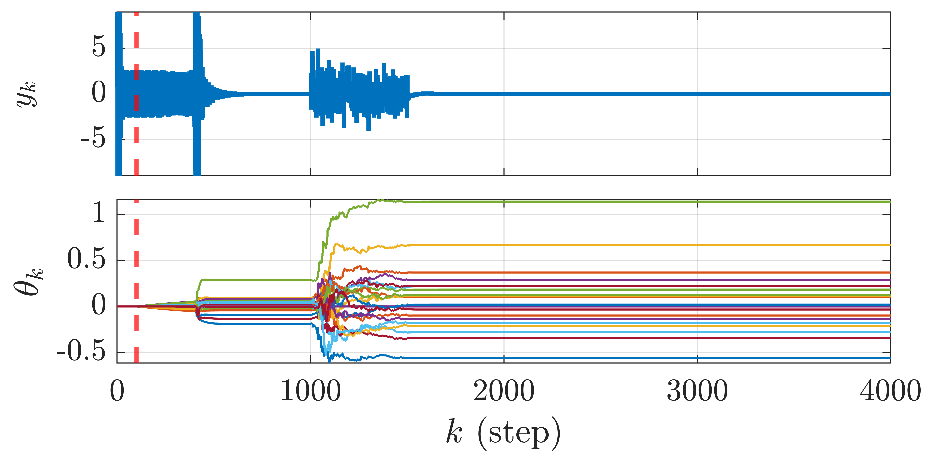}
    }
    \vspace{-2em}
    \caption{Example \ref{ex_1}: 
    DTL system output $y$ and the adaptive controller coefficients $\theta_k$ in the case where $x_0 = 1000B,$ and $v_k = v_{{\rm rand}, k}$. 
    The simulation transitions from open-loop operation to closed-loop operation at the step indicated by the vertical, dashed red line.
    }
    \label{fig:ex1_time_Vrand}
    \vspace{-2em}
\end{figure}
\begin{figure}[h!]
    \centering
    \vspace{0.5em}
    \resizebox{\columnwidth}{!}{%
    \includegraphics{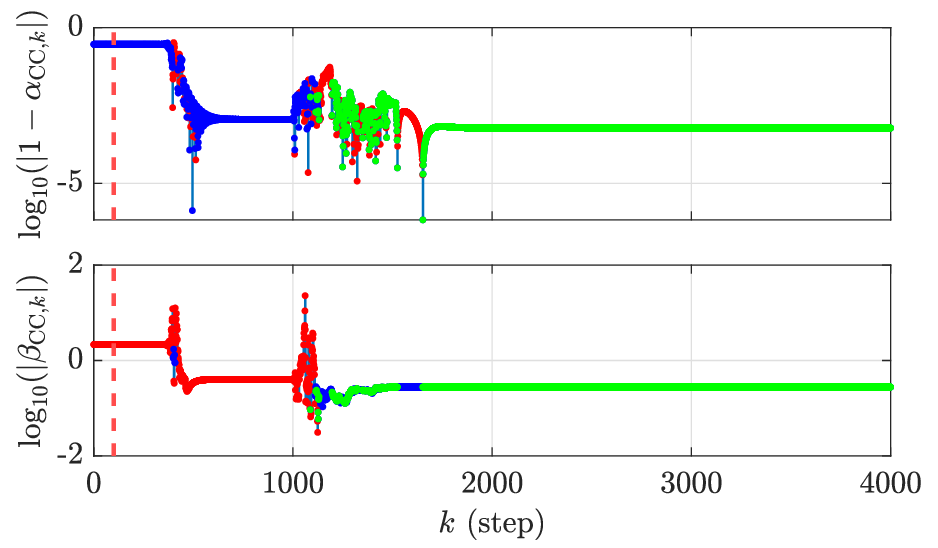}
    }
    \vspace{-2em}
    \caption{Example \ref{ex_1}:
    Evaluation of (CC1) and (CC2) in the case where $x_0 = 1000B,$ and $v_k = v_{{\rm rand}, k}$.  
    Values in red correspond to steps at which $1 - \alpha_{{\rm CC}, k}$ and  $\beta_{{\rm CC}, k}$ are nonpositive.
    Values in blue correspond to steps at which $1 - \alpha_{{\rm CC}, k}$ and  $\beta_{{\rm CC}, k}$ are positive.
    The closed-loop system satisfies (CC1) and (CC2) when the values in both plots are green.
    For this case, the circle criterion is satisfied for all  $k\ge 1700.$ 
    }
    \label{fig:ex1_circ_Vrand}
    \vspace{-1.3em}
\end{figure}
\begin{figure}[h!]
    \centering
    \resizebox{\columnwidth}{!}{%
    \includegraphics{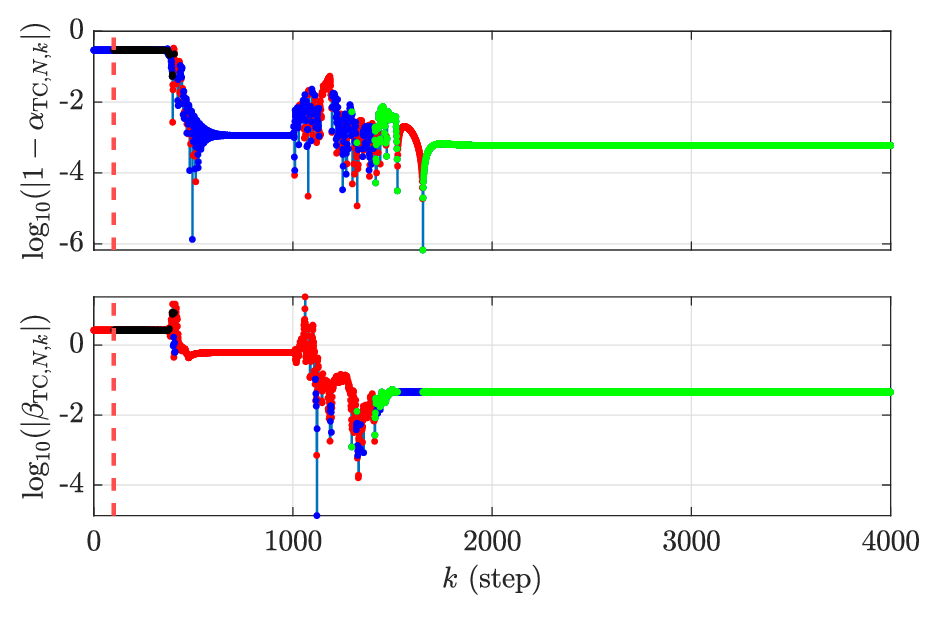}
    }
    \vspace{-2.5em}
    \caption{Example \ref{ex_1}:
    Evaluation of the Tsypkin criterion  in the case where $x_0 = 1000B,$ and $v_k = v_{{\rm rand}, k}$  with $N = 0.08.$ 
    Values not in black correspond to steps at which $\zeta_{1,k} \neq 0,$ $\zeta_{2,N,k} = n + \hat{n}p,$ and $\zeta_{3,N,k} > 0.$
    Values in red correspond to steps at which $1 - \alpha_{{\rm TC}, N, k}$ and $\beta_{{\rm TC}, N, k}$ are nonpositive.
    Values in blue correspond to steps at which $1 - \alpha_{{\rm TC}, N, k}$ and $\beta_{{\rm TC}, N, k}$ are positive.
    The closed-loop system satisfies (TC1), (TC2) and (TC3) when the values in both plots are green.
    For this case, the Tsypkin criterion is satisfied for all $k\ge 1700$.
    }
    \label{fig:ex1_tsyp_Vrand}
    \vspace{-1.95em}
\end{figure}

\begin{exam}\label{ex_1}
Let $G (\bfq) = (\bfq - 1)/(\bfq^2 - \bfq + 0.5)$ with minimal realization
\begin{equation}\label{eq:MR_ex_1}
    A = \begin{bmatrix} 1 & -0.5 \\ 1 & 0 \end{bmatrix}, \ B = \begin{bmatrix} 1 \\ 0 \end{bmatrix}, \ C = \begin{bmatrix} 1 & -1 \end{bmatrix},
\end{equation}
$x_0 = 1000 B,$ and $\gamma(y) = \tanh(y).$ 
The open-loop response of the DTL system and the nonlinearity $\gamma$ are shown in Figure \ref{fig:ex1_OL}.
Note, $\gamma$ is SB and DISB with sector bound $[0, 1].$

In the case where $v_k = 0,$ the results for $k\in [0, 1000]$ are shown in Figures \ref{fig:ex1_time}, \ref{fig:ex1_circ}, and \ref{fig:ex1_tsyp}.
Note that, although the output $y_k$ of the closed-loop DTL system converges to zero, the circle and Tsypkin criteria are not satisfied at $k = 1000.$
Figure \ref{fig:ex1_roa} shows, nevertheless, that the  closed-loop Lur'e system with the converged LTI linear dynamics $\tilde{G}_{1000}$ has a large region of attraction.

Next, for $v_k = v_{{\rm imp}, k},$
the DTL system output is shown in Figures \ref{fig:ex1_time_Vimp}, \ref{fig:ex1_circ_Vimp}, and \ref{fig:ex1_tsyp_Vimp}.
Note that the output $y_k$ of the closed-loop DTL system converges to zero, and the circle and Tsypkin criteria are satisfied at $k = 3000.$











Finally, for $v_k = v_{{\rm rand}, k},$
the DTL system output is shown in  Figures \ref{fig:ex1_time_Vrand}, \ref{fig:ex1_circ_Vrand}, and \ref{fig:ex1_tsyp_Vrand}.
Note that the output $y_k$ of the closed-loop DTL system converges to zero, and the circle and Tsypkin criteria are satisfied at $k = 2700.$
\hfill $\huge\diamond$

\end{exam}


\vspace{-1.1em}

\section{Conclusions}\label{sec:conclusion}
This paper analyzed the closed-loop stability of predictive cost adaptive control for output-feedback of a discrete-time, Lur'e system using absolute stability criteria.
Two absolute stability criteria for discrete-time systems were considered, namely, the circle criterion and the Tsypkin test.
Since the linear dynamics of the closed-loop Lur'e system are time-varying, these tests were applied instantaneously as a means for determining whether or not the adaptive controller is converging to a GAS system.
A numerical example showed that, under additional excitation, the circle and Tsypkin criteria are satisfied, suggesting that PCAC globally asymptotically stabilizes the DTL system. 
%




In this application of PCAC, linear-system identification is applied to a  nonlinear system that exhibits self-oscillations;  this behavior is not shared by the linearized system. 
Future research will thus focus on understanding the reasons for this effectiveness despite the modeling mismatch.

\vspace{-.1in}


\section*{Acknowledgments}

This research was supported by ONR under grant N00014-18-1-2211 and AFOSR under grant FA9550-20-1-0028.
The authors thank Wassim Haddad for helpful discussions on absolute stability.
%


\vspace{-.1in}

\appendix

\setcounter{equation}{0}
\renewcommand{\theequation}{\Alph{section}.\arabic{equation}}

\setcounter{theo}{0}
\renewcommand{\thetheo}{\Alph{section}.\arabic{theo}}

\setcounter{figure}{0}
\renewcommand{\thefigure}{\Alph{section}.\arabic{figure}}

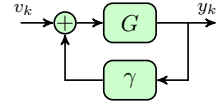
\begin{figure} [h!]
    \centering
    \vspace{1em}
    \resizebox{0.35\columnwidth}{!}{%
    \begin{tikzpicture}[>={stealth'}, line width = 0.25mm]

        \node [input, name=input] {};
		\node [sum, fill=green!20, right = 0.5cm of input] (sum1) {};
		\node[draw = white] at (sum1.center) {$+$};
		\node [smallblock, fill=green!20, rounded corners, right = 0.4cm of sum1, minimum height = 0.6cm, minimum width = 0.8cm] (system) {$G$};
		\node [smallblock, fill=green!20, rounded corners, below = 0.25cm of system, minimum height = 0.6cm, minimum width = 0.8cm] (satF) {$\gamma$};
		\draw [->] (input) -- node[name=usys, xshift = -0.2cm, yshift = 0.2cm] {\footnotesize$v_k$} (sum1.west);
		\draw[->] (sum1.east) -- (system.west);
		\node [output, right = 0.9cm of system] (output) {};
		\draw [->] (system) -- node [name=y,near end, xshift = -0.2cm]{} node [near end, above, xshift = 0.1cm] {\footnotesize$y_k$}(output);
		\draw [->] (y.center) |- (satF.east);
		\draw [->] (satF.west) -| (sum1.south);
    
    \end{tikzpicture}
    }  
    \caption{
    Discrete-time Lur'e (DTL) system composed of the positive feedback interconnection of $G$ and $\gamma.$
    }
    \label{fig:DTL_blk_diag}
    \vspace{-1.75em}
\end{figure}

Let $G(\bfq) \isdef C(\bfq I_n - A)^{-1}B$ be a strictly proper, discrete-time
transfer function with $n$th-order minimal realization $(A, B, C)$ and state $x_k \in \BBR^n$, let $v_k \in \BBR^m$ be an exogenous input, let $y_k\in\BBR^p$ be the output of $G,$ and let $\gamma\colon \BBR^p \to \BBR^m.$
Then, for all $k\ge0,$ the discrete-time Lur'e (DTL) system composed of the positive feedback interconnection of $G$ and $\gamma$ shown in Figure \ref{fig:DTL_blk_diag} has the closed-loop dynamics
\begin{align}
    x_{k+1} &= A x_k +  B (\gamma(y_k) + v_k), \label{xLureA}\\
    y_k &= C x_k.\label{yLureA}
\end{align}

\begin{defin} \label{def_sector_bounded}
Let $\gamma\colon \BBR^p \to \BBR^m$, and let $M_1 \in \BBR^{m \times p}$ and $M_2 \in \BBR^{m \times p}$ be such that $M_2 - M_1$ is positive definite.
Then, 
$\gamma$ is {\it sector-bounded (SB) with sector bound} [$M_1, M_2$] if, for all $y \in \BBR^p,$ $[\gamma(y) - M_1 y]^\rmT [\gamma(y) - M_2 y] \leq 0.$ 
\end{defin}

The following result is the circle criterion \cite[Thm 5.1]{haddad1994DT}. 


\begin{theo}\label{theo:AbsStabCC}
Assume that $G$ is strictly proper, 
let $M_1 \in \BBR^{m \times p}$ and $M_2 \in \BBR^{m \times p}$ be such that $M_2 - M_1$ is positive definite,
assume that $\gamma$ is SB with sector bound [$M_1, M_2$], 
define
$H(\bfq) \isdef [I_m - M_2 G(\bfq)] [I_m - M_1 G(\bfq)]^{-1},$
and assume that the following conditions are satisfied:
\small
\begin{itemize}
\item[] \hspace{-2em} (CC1) $\alpha_{\rm CC} \isdef {\rm spr} (H) < 1.$
\item[] \hspace{-2em} (CC2) $\beta_{\rm CC} \isdef \min_{\psi \in [0, \pi]} \left( \lambda_{\rm min} \left[ H (e^{\jmath \psi}) + H^\rmT (e^{-\jmath \psi}) \right] \right) > 0.$ 
\end{itemize}
\normalsize
Then, the zero solution of \eqref{xLureA}, \eqref{yLureA} is GAS.
\end{theo}

The following definition considers the case where $m = p.$

\begin{defin} \label{def_mono_sector_bounded}
Let $\gamma\colon \BBR^m \to \BBR^m$ be continuous, and let $M \in \BBR^{m \times m}$  be positive definite.
Then, $\gamma$ is {\it diagonal,  increasing, and sector-bounded (DISB) with sector bound} $[0, M]$ if the following conditions are satisfied:
\begin{enumerate}
%
%
\item For all $y \in \BBR^m,$  
$\gamma(y) = [\gamma_{(1)} (y_{(1)}) \cdots \gamma_{(m)} (y_{(m)}) ]^\rmT.$ 
\item For all $y \in \BBR^m,$  
$\gamma^\rmT(y) [\gamma(y) - M y] \leq 0.$
\item For all $i \in \{1, \ldots, m\}$ and distinct $\sigma, \hat{\sigma} \in \BBR,$ it follows that $(\gamma_{(i)} (\sigma) - \gamma_{(i)} (\hat{\sigma}))/(\sigma - \hat{\sigma}) > 0.$
\end{enumerate}
\end{defin}

Next is the Tsypkin criterion \cite[Thm 3.1]{kapila1996}.


%
\begin{theo}\label{theo:AbsStabTC}
Assume  $m = p,$ $G$ is strictly proper,  $\gamma$ is {\it DISB} with sector bound $[0, M],$ let $N \isdef {\rm diag} (N_1, \ldots, N_m)$ be positive definite, define
$L_N(\bfq) \isdef M^{-1} - [I_m + (1 - \bfq^{-1}) N] G(\bfq),$
and assume  the following conditions hold:
\small
\begin{itemize}
\item[] \hspace{-2em} (TC1) ${\rm det} (C A^{-1} B) \neq 0,$ $(A, C + N C - N C A^{-1})$ is observable, and $\lim_{\bfz \to \infty} [L_N (\bfz) + L_N(\bfz)^\rmT] >0.$
\item[] \hspace{-2em} (TC2) $\alpha_{{\rm TC}, N} \isdef {\rm spr} (L_N)< 1.$
\item[] \hspace{-2em} (TC3) $\beta_{{\rm TC}, N} \isdef \min_{\psi \in [0, \pi]} ( \lambda_{\rm min} [ L_N (e^{\jmath \psi}) + L_N^\rmT (e^{-\jmath \psi}) ] ) > 0.$ 
\end{itemize}
\normalsize
Then, the zero solution of \eqref{xLureA}, \eqref{yLureA} is GAS.
\end{theo}

\vspace{-.05in}

\bibliographystyle{IEEEtran}
\bibliography{IEEEabrv,bib_paper}

\end{document}